\def \SAIT #1 #2 {{\em Mem.\ Soc.\ Astron.\ It.\/} {\bf #1}, #2}
\def \MESS #1 #2 {{\em The Messenger\/} {\bf #1}, #2}
\def \ASTRNACH #1 #2 {{\em Astron. Nach.\/} {\bf #1}, #2}
\def \AAP #1 #2 {{\em Astron. Astrophys.\/} {\bf #1}, #2}
\def \AAL #1 #2 {{\em Astron. Astrophys. Lett.\/} {\bf #1}, L#2}
\def \AAR #1 #2 {{\em Astron. Astrophys. Rev.\/} {\bf #1}, #2}
\def \AAS #1 #2 {{\em Astron. Astrophys. Suppl. Ser.\/} {\bf #1}, #2}
\def \AJ #1 #2 {{\em Astron. J.\/} {\bf #1}, #2}
\def \ANNREV #1 #2 {{\em Ann. Rev. Astron. Astrophys.\/} {\bf #1}, #2}
\def \APJ #1 #2 {{\em Astrophys. J.\/} {\bf #1}, #2}
\def \APJL #1 #2 {{\em Astrophys. J. Lett.\/} {\bf #1}, L#2}
\def \APJS #1 #2 {{\em Astrophys. J. Suppl.\/} {\bf #1}, #2}
\def \APSS #1 #2 {{\em Astrophys. Space Sci.\/} {\bf #1}, #2}
\def \ASR #1 #2 {{\em Adv. Space Res.\/} {\bf #1}, #2}
\def \BAIC #1 #2 {{\em Bull. Astron. Inst. Czechosl.\/} {\bf #1}, #2}
\def \JSQRT #1 #2 {{\em J. Quant. Spectrosc. Radiat. Transfer\/} {\bf #1}, #2}
\def \MN #1 #2 {{\em Mon. Not. R. Astr. Soc.\/} {\bf #1}, #2}
\def \MEM #1 #2 {{\em Mem. R. Astr. Soc.\/} {\bf #1}, #2}
\def \PLR #1 #2 {{\em Phys. Lett. Rev.\/} {\bf #1}, #2}
\def \PASJ #1 #2 {{\em Publ. Astron. Soc. Japan\/} {\bf #1}, #2}
\def \PASP #1 #2 {{\em Publ. Astr. Soc. Pacific\/} {\bf #1}, #2}
\def \NAT #1 #2 {{\em Nature\/} {\bf #1}, #2}

\documentstyle{memsait}
\input epsf.sty
\begin{opening}
\title{Thermally Pulsing AGB Models of Intermediate Mass Stars}
\author{Oscar Straniero$^1$, Marco Limongi$^2$, Alessandro Chieffi$^3$,\\
 Inma Dominguez$^4$, Maurizio Busso$^5$, Roberto Gallino $^6$}
\institute{
 $^1$Osservatorio Astronomico di Collurania, 64100 Teramo, Italy\\
 $^2$Osservatorio di Roma, Via Osservatorio 2, 00040 Monte Porzio (RM), Italy\\
 $^3$Istituto di Astrofisica Spaziale (CNR), Roma, Italy\\
 $^4$Dpto. de F\'{\i}sica Te\'orica y del Cosmos, Universidad de Granada, 18071 Granada, Spain\\
 $^5$Osservatorio Astronomico di Torino, 10025 Torino, Italy\\
 $^6$Dipartimento di Fisica Generale, Universita' di Torino, V. P. Giuria 1, 10125 Torino, Italy}
\date{} 
\end{opening}

\begin{document}

\oddpagefooter{}{}{} 
\evenpagefooter{}{}{} 
\ 
\bigskip

\begin{abstract}
We present a set of models of AGB stars with initial mass larger than
5 $M_{\odot}$ as obtained with the FRANEC code. It includes model of 
Z=0.02 and Z=0.001, with and without mass loss.  
\end{abstract}

\section{Introduction}
In the last years we have computed several sequences of AGB models
of intermediate mass stars with the purpose
to investigate the evolutionary properties and the related nucleosynthesis
of these class of objects. In this paper we have collected all these models in order to
summarize our main findings. In Table 1 we report a list of the computed sequences. Some 
features of the models are reported too. Let us remind that the models 
presented here have been obtained with the same version
of the FRANEC code used in our previous computation of low
mass AGB stars (Straniero et al. 1997). 

\vspace{1cm} 
\centerline{\bf Tab. 1}
\begin{table}[h]
\hspace{2.5cm} 
\begin{tabular}{|l|c|c|c|c|c|}
\hline
\hline
Mass       &5       &6      &7       &5       &5      \\
Z          &0.02    &0.02   &0.02    &0.02    &0.001  \\
Y          &0.28    &0.28   &0.28    &0.28    &0.23   \\
Mass loss  &no      &no     &3       &10      &no     \\
N. of TPs  &58      &51     &59      &23      &40     \\
Final Mass &5       &6      &6.42    &2.48    &5      \\
Final C/O  &0.9     &0.52   &0.34    &0.6     &1.94   \\
$T_{BCE}$  &46      &69     &82      &17      &73     \\
$T_{CSH}$  &361     &356    &360     &345     &370    \\
${\Delta}M_{DU}$    &1.3E-3 &6.5E-4  &2.2E-4  &1.4E-3  &1.2E-3   \\
\hline
\hline
\end{tabular}
\end{table}

The ten rows in Table 1 report
respectively: the initial mass (ZAMS mass) of each sequence, the initial metallicity, 
the initial helium, the adopted mass loss rate ("no" means no mass loss, while numbers 
indicates the value of the ${\eta}$ parameter in the Reimers formula), the numbers of computed
thermal pulses, the final (last computed) mass, the final
(last computed) carbon over oxygen ratio, the maximum temperature (in $10^6$ K)
at the base of the convective envelope 
(generally it coincides with the one obtained in the last computed interpulse
except in the case of the 5 M$_{\odot}$ $\eta=10$), the maximum temperature
 (in $10^6$ K) at the base of the He convective 
shell and, finally, the last computed value of the amount of mass (in $M_{\odot}$) 
dredged from the He-core. Note that all the sequences (except the one of the 
5 $M_{\odot}$ 
with ${\eta}=10$) were arbitrarily stopped after about 40-50 TPs. The 
case of the 5 $M_{\odot}$ with ${\eta}=10$ will be discussed below.
\begin{figure}
\epsfysize=6cm 
\hspace{3.5cm}\epsfbox{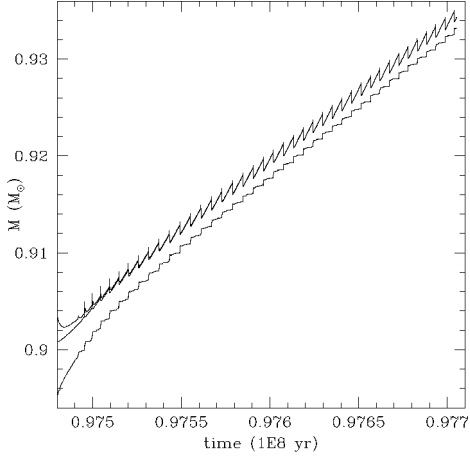} 
\caption[h]{Evolution of the inner edge of the convective envelope together with the 
location of the H burning shell and  the location of the He burning shell, for
the 5 $M_{\odot}$ Z=0.001.}
\end{figure}

\section{The III dredge-up}
At variance with low mass stars, the III dredge-up (TDU) occurs rather early in our
thermally pulsing intermediate mass models. For Z=0.02, 
the first evident episode is found after the 4th, the 5th and the 7th thermal pulse
in the 5, 6 and 7 $M_{\odot}$ respectively, whereas at Z=0.001 the
5 $M_{\odot}$ experiences a TDU just after 3 TPs.
The positions of the H and He burning shells as well as the position of the base of the 
convective envelope are reported in Figure 1, for the 5 $M_{\odot}$ Z=0.001.
We found that the penetration of the convective
envelope into the He core decreases when the mass increases (see the last row in Table 1).
In addition the extension (in mass) of the region between the two burning shells is
lower for the more massive models.  
These occurrences are probably due to the connection between the strength of
the pulse and the efficiency of the
Hydrogen burning shell (HBS). In fact the ignition point of the He burning shell
(HeBS) in the temperature/density plane depends on the H burning rate.
The larger this rate the lower the density and the larger the temperature of
the He shell at the moment of the re-ignition. Then, less work must be done by the 
He burning to expand the lighter layers above it and, in turn, a weaker pulse and a
smaller TDU occur. In the last computed pulse of the 5 $M_{\odot}$ 
the 3${\alpha}$ luminosity peak 
exceeds $10^8$ $L_{\odot}$, but it drops to $6\cdot10^7$ and $2\cdot10^7$ $L_{\odot}$ in the
6 and 7 $M_{\odot}$ models, respectively (see also Figure 2 and 3).
\begin{figure}
\epsfysize=6cm 
\hspace{3.5cm}\epsfbox{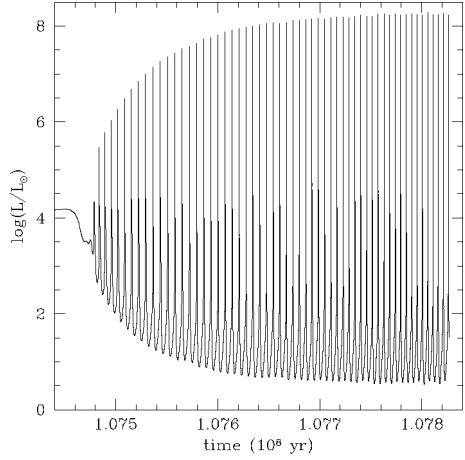} 
\caption[h]{Variation of the He burning luminosity for the 
5 $M_{\odot}$ (no mass loss) Z=0.02.}
\end{figure}
\begin{figure}
\epsfysize=6cm 
\hspace{3.5cm}\epsfbox{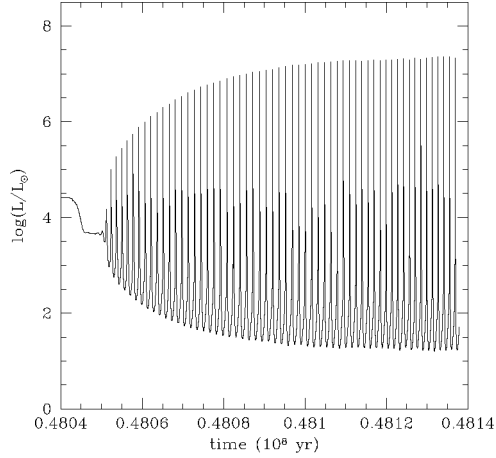} 
\caption[h]{Variation of the He burning luminosity for the
7 $M_{\odot}$ Z=0.02.}
\end{figure}

The more massive models have 
a more efficient HBS because of the larger core mass and the deeper penetration of the
H rich convective envelope into the burning region (see next section).
We have tested such a connection between H burning efficiency, pulse strength and TDU,
by artificially reducing (or increasing) the rate of the $^{14}$N(p,$\gamma$)$^{15}$O), 
which is the bottleneck of the CNO cycle. This test confirms our hypothesis and allow us to
conclude that any input
physics, which could alter the rate of the H burning, have a direct influence on the 
features of the thermal pulse, in particular its strength and dredge-up.  

\begin{figure}
\epsfysize=6cm 
\hspace{3.5cm}\epsfbox{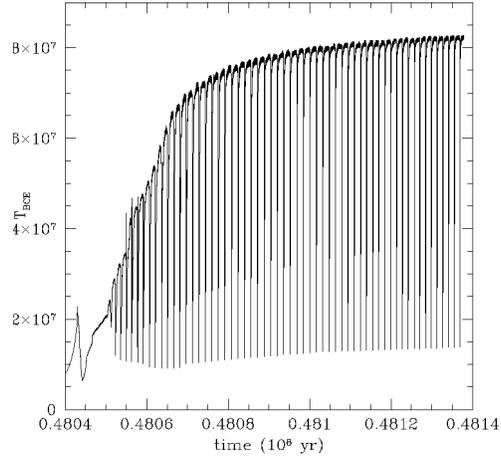} 
\caption[h]{Temperature at the base of the convective envelope for the
7 $M_{\odot}$.}
\end{figure}
\begin{figure}
\epsfysize=6cm 
\hspace{3.5cm}\epsfbox{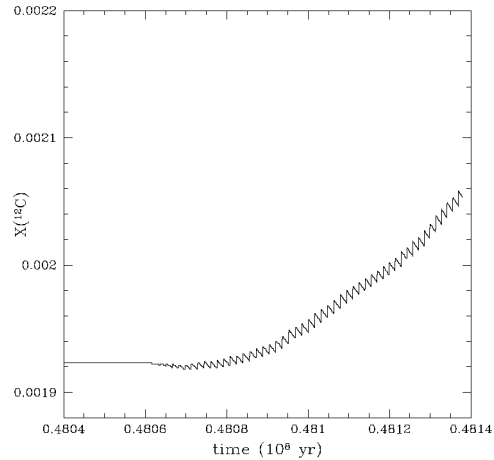} 
\caption[h]{Variation of the surface mass fraction of carbon in the
7 $M_{\odot}$.}
\end{figure}

\section{Hot Bottom Burning}
The evolution of the temperature at the bottom of the convective envelope (TBCE) for our
7 $M_{\odot}$ model of solar chemical composition is reported in Figure 4. 
We can see that this temperature rapidly approaches $8\cdot10^7$ K.
Similar values were obtained  by Bl\"ocker (1995) for a 7 $M_{\odot}$ and by Lattanzio et al. (1996)
for a 6 $M_{\odot}$.
As shown in Figure 5 these temperatures are so large that most of the carbon dredged up
from the He shell by the TDU is converted into nitrogen during the interpulse.
A similar situation is found for the 
6 $M_{\odot}$, although the maximum TBCE is slightly lower (about $7\cdot10^8$ K). 
However, the $^{12}$C(p,${\gamma}$)$^{13}$N reaction is almost inactive
at the base of the convective envelope
in the 5 $M_{\odot}$. In this case the maximum TBCE never exceeds $5\cdot10^7$ K. 

As already noted by Lattanzio
et al. (1996) HBB is strongly limited by mass loss. In the extreme case of the  
5 $M_{\odot}$ ${\eta}=10$ the maximum TBCE does never reach $2\cdot10^7$ K.

At lower metallicity, the thickness of the HBS is larger and the penetration
of the convective
envelope into the region of the CNO burning is favored by
a less steep entropy barrier. In our 5 $M_{\odot}$ Z=0.001, 
we found a bottom temperature as large as $7\cdot10^7$ K and a significant HBB. 
(see Figure 6 and 7). This is in agreement with the result of Lattanzio \& Forestini (1999). 
\begin{figure}
\epsfysize=6cm 
\hspace{3.5cm}\epsfbox{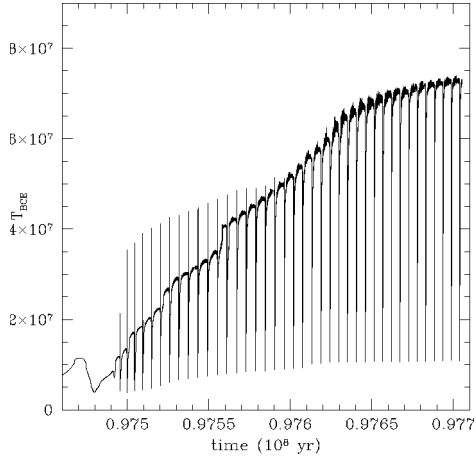} 
\caption[h]{Temperature at the base of the convective envelope for the
5 $M_{\odot}$ Z=0.001.}
\end{figure}
\begin{figure}
\epsfysize=6cm 
\hspace{3.5cm}\epsfbox{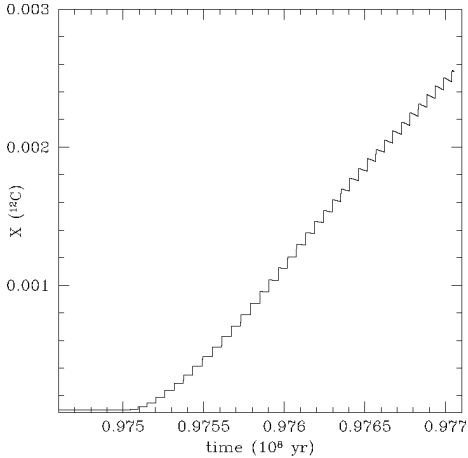} 
\caption[h]{Variation of the surface mass fraction of carbon in the
5 $M_{\odot}$ Z=0.001.}
\end{figure}

As firstly found by Bl\"ocker \& Sch\"onberner (1991) in numerical computations of AGB stars,
the classical core mass/luminosity relation
(Paczynski, 1975, Iben \& Renzini 1983) cannot apply to stars which experience  
HBB. Although
we found a certain deviation from the classical relation, 
the luminosity of our more massive models, which are close to $M_{up}$ (i.e. the minimum mass
for the degenerate carbon ignition), never exceeds
the observed magnitude of the AGB tip ($M_{bol}\sim-7$, see e.g. Wood et al. 1992).
\begin{figure}
\epsfysize=6cm 
\hspace{3.5cm}\epsfbox{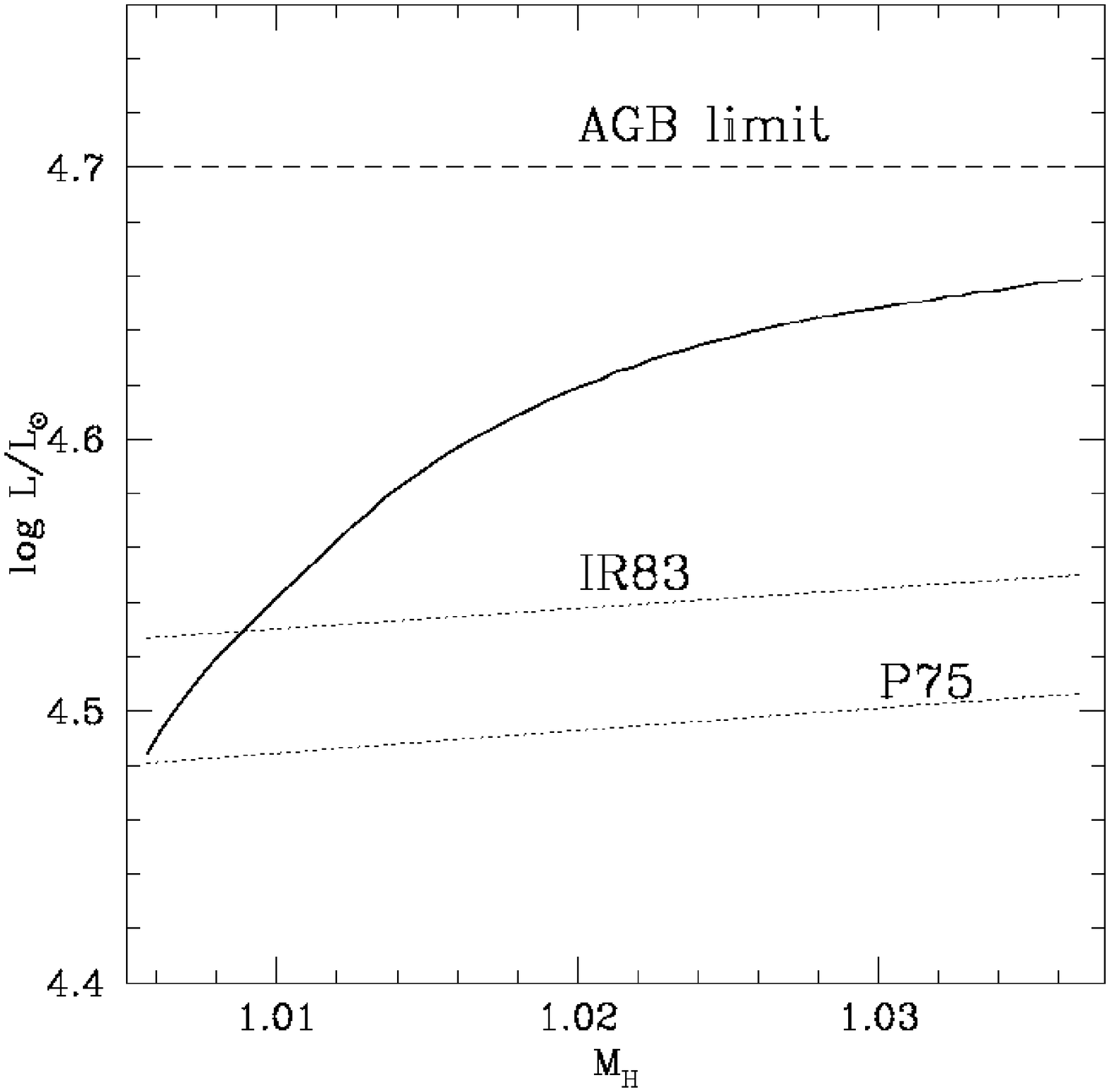} 
\caption[h]{The core mass/luminosity relation for the 7 $M_{\odot}$.
Solid line represent the maximum luminosity during the interpulse. The two 
dashed line indicate the classical relation by Paczynski (1975, P75) and that by
Iben \& Renzini (1983, IR83). The observed AGB limit (Wood et al., 1992) is reported too}
\end{figure}
In Figure 8 we report the
luminosity versus the core mass for our model of 7 $M_{\odot}$. Note how the 
luminosity of these models asymptotically approaches the observed AGB limit.

\section{Neutron source and s-process nucleosynthesis}
At the time of the TDU, if
a certain amount of protons are diffused below the 
convective envelope into the top layer of the Helium/carbon rich 
region, a tiny $^{13}$C pocket forms. Later on, as already found for low mass stars
(Straniero et al. 1995, Straniero et al. 1997, Herwig et al. 1997),
when the temperature in that pocket rises up to 90-100 $10^6$ K,
the $^{13}$C is fully destroyed  by ${\alpha}$ capture during the interpulse.
So neutrons are released and the heavy {\it s}-elements production can take place. 
In such a
case the typical neutron density is of the order of $10^7-10^8$ $n/cm^3$.
In our models with $M\ge5$ $M_{\odot}$, a further neutron source operates during the
convective thermal pulse. In fact, as shown in Table 1, the temperature at the base
of the He convective shell
reaches $3.5\cdot10^8$ K. In such condition, the substantial activation of the 
$^{22}$Ne(${\alpha}$,n)$^{25}$Mg reaction provides an important contribution to
the {\it s}-process
nucleosynthesis. This confirms the old prescription of Iben (1975a,b).
The neutron density during the TP is definitely greater than in the case of the $^{13}$C
burning.
We found ${\rho}_n\sim10^{11}$ $n/cm^3$. Some details of the  nucleosynthesis in
intermediate mass stars can be found in Vaglio et al. (1999).

\section{Termination of the AGB}

The search of the physical mechanism responsible for the formation of a Planetary Nebula is
a longstanding problem in modern astrophysics. This subject is obviously related 
to the comprehension of the final stage of the AGB evolution. In model computations
it is currently assumed that, at a certain point along the AGB,  mass loss
abruptly grows (up to $10^{-5}-10^{-4} M_{\odot}/yr$), so that an envelope ejection
is simulated (see e.g. Vassiliadis \& Wood, 1993; Bl\"ocker, 1995;
Forestini \& Charbonnel, 1997).
However, the correct evaluation of the masses of planetary nebulae and  
those of their nuclei,
as well as the description of the AGB termination will depend on the particular 
parameterization of this superwind regime. 

\begin{figure}
\epsfysize=6cm 
\hspace{3.5cm}\epsfbox{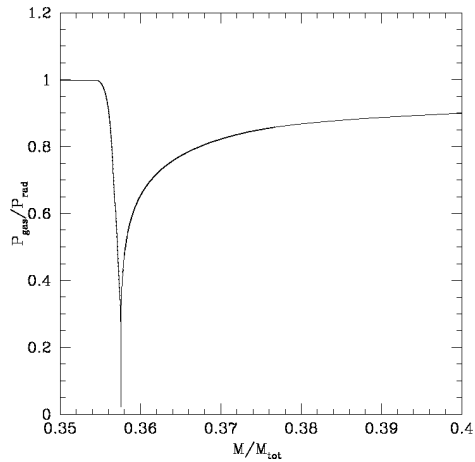} 
\caption[h]{The ratio of the gas pressure to the radiation pressure
in the last computed model of the 5 $M_{\odot}$ $\eta=10$. The upper part of the
He core and the most internal
region of the convective envelope are shown.}
\end{figure}

In a recent paper Sweigart (1998) found that in mass losing models 
the ratio ($\beta$)
of the radiation pressure to the gas pressure drops abruptly in a region located
just above the  H/He discontinuity. 
This occurs just after each thermal pulse. 
When the envelope mass is reduced enough (how much is depending on the core mass 
and metallicity), $\beta$ becomes practically 0 and the local stellar luminosity
exceeds the Eddington 
luminosity. Then, an instability, which could drive the 
envelope ejection, settles on.
Something similar was also reported by Wood \& Faulkner (1986).   

We confirm the Sweigart's finding.  
In our 5 $M_{\odot}$ ${\eta}=10$ the minimum value of $\beta$ is attained 
during the TDU. At the beginning of the 
TP-AGB phase, the value of this
minimum is about 0.3-0.4, but it decreases from one pulse to the next
as the envelope mass is reduced by the mass loss. After 23 thermal pulses, when the 
residual total mass is 2.48 $M_{\odot}$ and the core mass is 0.89 $M_{\odot}$,
$\beta$ goes to 0 and our hydrostatic code cannot go ahead (see Figure 9).
We found the same situation in a 2.5 $M_{\odot}$, Z=0.006  and ${\eta}=2$.
In such a case the final (last computed) mass is 0.81 $M_{\odot}$
and the core mass is 0.67 $M_{\odot}$. 

Such an occurrence have a quite simple explanation. 
As a consequence of the thermal pulse, an expansion, starting 
from the He shell, propagates toward the surface. When this expansion 
reaches the base of H rich envelope, the local temperature decreases
below $10^6$ K. Then the drop of $\beta$ is determined by the well known bump
in the metal opacity around log {\it T}=5.3 (see e.g. Iglesias, Roger, Wilson, 1992).
which implies a decrease of the local Eddington luminosity. 
Thus a sort of void forms between the core and the envelope
(see Figure 9).
In this thin layer the stellar structure cannot react, as usually occurs,
to an increase of the local temperature by expanding 
the gas and reducing the local pressure. Then, any small
perturbation will inevitably grow. We cannot say if this instability
can indefinitely grow, but we believe that this phenomenon could play a pivotal role
in the AGB termination. This problem deserves a further investigation,
possibly by means of an hydrodynamic code.
Let us finally note that the TDU increases the metal content of the envelope,
and, in turn, the opacity bump will increase. So the larger the dredge-up the  
deeper the $\beta$ drop is. In the present computations the final C/O ratio in the envelope
of the 5 $M_{\odot}$ is 0.6, whereas in the case of the 2.5 $M_{\odot}$ we obtain
a carbon star well before the end of the sequence.

\acknowledgements
This work was partially supported by the MURST Italian grant Cofin98,
by the MEC Spanish grant PB96-1428, by the Andalusian grant FQM-108 and 
 it is part of the ITALY-SPAIN
integrated action (MURST-MEC agreement) HI1998-0095.

\end{document}